# Seasonal and Periodic Patterns in US COVID-19 Mortality using the Variable Bandpass Periodic Block Bootstrap

Edward Valachovic*

evalachovic@albany.edu

Ekaterina Shishova*

*Department of Epidemiology and Biostatistics, School of Public Health, University at Albany, State University of New York, One University Place, Rensselaer, NY 12144

**Abstract**

Since the emergence of the SARS-CoV-2 virus, research into the existence, extent, and pattern of seasonality has been of the highest importance for public health preparation. This study uses a novel bandpass bootstrap approach called the Variable Bandpass Periodic Block Bootstrap (VBPBB) to investigate the periodically correlated (PC) components including seasonality within US COVID-19 mortality. Bootstrapping to produce confidence intervals (CI) for periodic characteristics such as the seasonal mean requires preservation of the PC component's correlation structure during resampling. While existing bootstrap methods can preserve the PC component correlation structure, filtration of that PC component's frequency from interference is critical to bootstrap the PC component's characteristics accurately and efficiently. The VBPBB filters the PC time series to reduce interference from other components such as noise. This greatly reduces bootstrapped CI size and outperforms the statistical power and accuracy of other methods when estimating the periodic mean sampling distribution. VBPBB analysis of US COVID-19 mortality PC components are provided and compared against alternative bootstrapping methods. These results reveal crucial evidence supporting the presence of a seasonal PC pattern and existence of additional PC components, their timing, and CIs for their effect which will aid prediction and preparation for future COVID-19 responses.

## 1. Introduction

In 2020 the SARS-CoV-2 virus, also known as COVID-19, became a significant global health crisis (World Health Organization, 2023). As the virus has transitioned over time from a global pandemic to a persistent endemic there is strong interest in any potential seasonal component of the disease (Rayan, 2021; Liu et al., 2021; Merow and Urban, 2020; Choi et al., 2021). This has become a subject of frequent and ongoing research (Wiemken et al., 2023; Kronfeld-Schor et al., 2021, Gavenčiak et al., 2022, Kumar et al. 2021). Better evidence and characterization of the seasonality of COVID-19 can prove a useful tool for public health planning, preparedness, allocation of resources, communication, and a directed response. With more than three years of data there is an increasing opportunity to research the periodically correlated (PC) time series components such as the seasonal pattern of COVID-19. This research investigates the seasonal PC component of COVID-19 mortality in the United States using a novel non-parametric resampling method called the Variable Bandpass Periodic Block Bootstrap (VBPBB) producing confidence intervals for the seasonal and other periodic means (Valachovic, 2024).

A brief timeline of COVID-19 can be found from the Center for Disease Control and Prevention (CDC) Museum COVID-19 Timeline (2023) and begins in December 2019 with what would become known as the SARS-CoV-2 virus emerging in Wuhan, in China's Hubei Province. By January 14, 2020 the World Health Organization (WHO) has the first evidence of possible human-to-human transmission of the SARS-CoV-2 virus. On January 20, 2020 CDC reports the first laboratory-confirmed case in the U.S. Spreading rapidly, by March 11, 2020, with in excess of 118,000 cases in 114 countries and 4,291 deaths, the WHO declares COVID-19 a pandemic. In the US, a nationwide emergency declaration is issued on March 13, 2020. Towards the present, the WHO ends the COVID-19 global health emergency on May 5, 2023, and by May 11, 2023 the US COVID-19 public health emergency declaration ends. During this time there are approximately 6.8 million deaths globally from COVID-19. As of July 1, 2023 there are approximately 1.135 million recorded deaths in the US from COVID-19. Despite the end of the emergency declarations, in July 2023 there are approximately 8000 US weekly hospital admissions, 1% of Emergency Department visits are diagnosed with COVID and there are approximately 500 weekly deaths according to the National Center for Health Statistics (2023).

The process of random sampling with replacement, or bootstrapping, from sampled data resulting in independent draws that form a resample of the same length as the original data set was first detailed by Efron (1979). Bootstrapping is a useful process to estimate the sampling distribution of statistics such as the mean. Bootstrapping a large number, $B$, of resamples replicates the sampling variability for a desired statistic. Calculating the $(1 - \alpha/2)$ and $(\alpha/2)$ quantiles of the $B$ bootstrapped statistic values produces a $100 * (1 - \alpha)\%$ confidence interval (CI) for the statistic. Time series data behave differently than typical sample data since successive data points, or observations, are ordered in time. More on time series including definitions, notation, and examples can be found in Wei (1989). Often, ordered data points may be correlated with prior observations in the time series. Consequently, independently resampling data points from the dataset to form a new ordered resample will not replicate any correlations between a given and prior data points within the original time series. The correlation between successive time series data points is destroyed. Special bootstrapping methods are needed to prevent the destruction of correlation structures within time series data. Block bootstrapping is a class of methods which attempt to replicate and preserve correlations in spatio-temporal data. Block bootstrapping often involves a general strategy of splitting the time series into blocks and then randomly sampling the blocks to form the resamples. An example of this is the Moving Block Bootstrap first introduced by Kunsch (1989). This can help replicate the correlation structure between given data points and some prior observations within the limits of the block size. Clearly, correlations between data points separated by more than the length of the bootstrapped blocks will have correlation structures destroyed.

When time series data are cyclic, cyclostationary, or periodic in nature such as seasonality when referring to the annual cycle, they possess a PC component. A PC component with a given period, $p$, or the corresponding reciprocal of the period, $1/p$, called the frequency, exhibits strong correlations between data points that are $kp$ time points separated, where $k \in \mathbb{Z}$ is an integer multiple. Block bootstrapping has difficulty reproducing the correlation structure of PC time series. As previously described, block bootstrapping with any block lengths less than the given period, $p$, will eliminate these correlations. Additionally, bootstrapping with block lengths greater than the given period, $p$, will preserve the correlation within each block; however, there is no reason randomly sampled adjacent blocks will retain synchronicity in the periodic correlation structure when forming the resample. For example, there is no guarantee the first data point of a resampled block to be the next step in each cycle of period $p$ that follows the last data point of the prior resampled block. For this reason, many block bootstrapping strategies are not well suited for PC time series.

A seasonal block bootstrap was proposed by Politis (2001) where the block length is restricted to be a multiple of the period $p$. Regardless of whichever step in the cycle a block begins, all other blocks in the resample will likewise start and end with a common step in the cycle of period $p$. This strategy will preserve correlation structures between data points that are $p$ points removed from each other in time series data. Some other block bootstrapping methods for PC time series include those presented by Chan et al. (2004) and the general seasonal block bootstrap (GSBB) by Dudek et al. (2014). While the block bootstrap methods described above, which can collectively be referred to as periodic block bootstrap (PBB) methods, attempt to preserve the PC component correlation structure, these strategies also bootstrap any additional interfering components including the noise present in the time series. Unsurprisingly, even modest levels of interference can inflate bootstrapped confidence interval sizes for measures such as the PC component periodic mean beyond the stated confidence interval level. In the context of COVID-19 mortality rates, a time series beset by instability such as an accelerating emergence, large interventions such as emergency declarations, lockdowns, vaccine introductions, booster development, reopening, and changes in trend, this interference can have a severe detrimental effect on producing CI for the periodic mean of a PC component.

This work uses the novel VBPBB which by design is highly resistant to the instabilities observed in COVID-19 data and outperforms other methods given that interference (Valachovic, 2024). VBPBB is less sensitive to the presence of shocks, interventions, potential long-term trends, noise, and outliers, making it better suited for the current application to COVID-19. The VBPBB applies a bandpass filter around a PC component frequency, passing or preserving time series variation that occurs at frequencies equal to or near the component frequency of interest. Variations at frequencies outside of the band are attenuated. The bandpass filter produces a time series that is primarily composed of only the PC component. The VBPBB then block bootstraps the PC component time series to preserve the correlation structure by selecting a block size equal to the PC component period. VBPBB yields smaller CI sizes for the periodic mean than PBB methods such as the GSBB. In this study, we address only the broad question of periodic patterns in the national statistics of COVID-19 mortality, but the approach is easily translated to other national COVID-19 patterns such as diagnoses or admissions, and at all stratified levels from national, state, county, or hospital level data, or to other epidemiological questions altogether such as the effect of COVID-19 response on other respiratory illness seasonality.

While the seasonal component of COVID-19 is of primary interest, a PC component may not behave as a perfect sinusoidal wave, strictly operating at one principal period or frequency. If a PC component of a period $p$ or frequency $1/p$, called the fundamental frequency, has a pattern other than a sinusoidal wave, it often has variations that occur at positive integer multiples of the period, $kp$ for $k \in \mathbb{Z}^+$, or frequency, $1/kp$, called the $k^{th}$ harmonics of the fundamental frequency. For this reason, this study also investigates the harmonics in addition to the seasonal pattern, and other PC components of United States COVID-19 mortality. Using both the PBB and VBPBB, 95% CI bands formed by the 95% CI for the periodic mean mortality at each time point throughout the period of the study. CI size is compared between the methods for the periodic means and significant PC components are identified and quantified.

## 2. Methods

### 2.1 Data Sources and Analysis

United States COVID-19 national mortality is recorded daily. This is compiled from the CDC, state, and local public health agencies and the time series is available at USAFacts (2023) and also from the National Center for Health Statistics (2023). The time series runs from January 22, 2020 through July 1, 2023. Mortality rates are given per hundred thousand after adjusting for population with data available from the United States Census Bureau (2023). The US COVID-19 mortality rate time series is seen in Figure 1. Analysis is performed in R version 4.1.1 statistical software (R Core Team, 2013). The primary PC component of interest is the annual or seasonal component, or variation in US COVID-19 mortality occurring in a 365-day pattern. As described above, this pattern may have harmonics, so secondary PC components of interest occurring at 365/2, 365/3, etc. are also included in this study. Additionally, the US COVID-19 mortality time series shows evidence of a weekly or 7-day variation, almost certainly resulting from human activities and interventions such as administrative recording practices across the states and local health agencies. Still, to better understand the periodic patterns, the 7-day PC component and its harmonics are also included in this study. Finally, this study investigated other potential PC components, such as Bi-weekly, Monthly, and other potential components, but there was no significant evidence of changes in the mean at these periods in US COVID-19 mortality, so they are excluded.

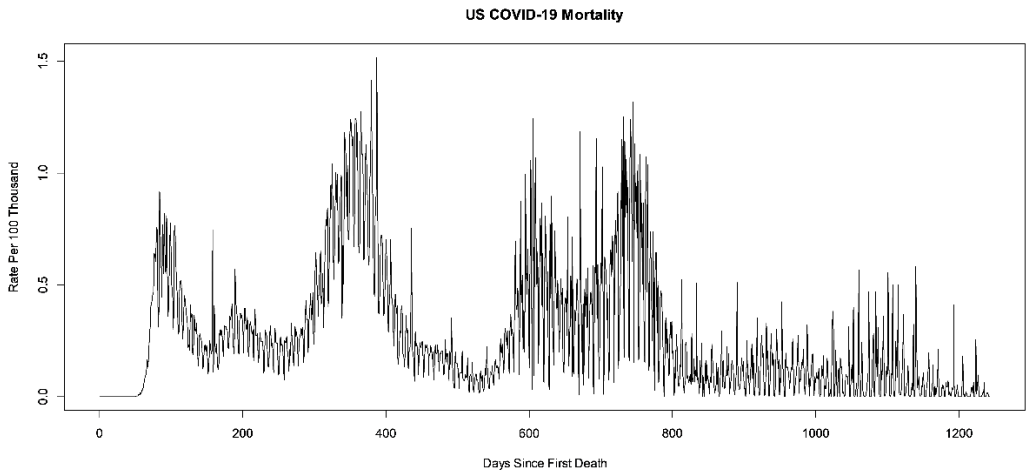

**Figure 1**: US COVID-19 mortality rate time series.

**2.2 PBB Approach**

For comparison, this work uses a PBB approach similar to the GSBB of Dudek et al. (2014). This approach applies the PBB without first bandpass filtering the PC components, the key difference with the VBPBB. The PBB block bootstraps the original COVID-19 mortality time series to preserve each PC component correlation structure first by fixing block size to $p$, the period of the target PC component. For a given period, $p$, for a time series of $n$ observations, the block bootstrap is created by forming $p$ exclusive and exhaustive subsets, each composed of one of each of the first $p$ observations and the integer multiples of that observation from the time series. Each resample is formed by independently, with replacement, randomly selecting from the $p$ subsets in sequence, repeated until $n$ observations are selected. This forms one bootstrap resample. This process is repeated for a large number, $B$, of resamples. Here, there are $B = 10000$ bootstrap resamples generated, each a time series of the same length as the original. The 0.975 and 0.025 quantiles of the *B* bootstrapped time series resamples are calculated to produce a 95% CI for the periodic mean at each time point. This forms a PBB 95% CI band for the periodic mean across the time interval.

## 2.3 VBPBB Approach

The objective of the VBPBB bootstrap approach is to resample a PC time series component alone to preserve the correlation structure of that PC component, while not resampling the other unrelated components such as noise, a linear trend, or discontinuity that will unnecessarily increase bootstrap variability. This is a weakness of the PBB approach. The VBPBB remedy is not found in the time domain, where components interact, but rather in the frequency domain. A mathematical fact is that components of different frequencies are uncorrelated. The VBPBB bandpass filters a PC time series according to its spectral density and separates a narrow band of frequencies including the PC component frequency from all others, creating a new PC time series composed primarily of the component PC of interest while preserving the PC component correlation structure. The new component PC time series, rather than the original time series, is block bootstrapped with the same design as described in the PBB approach above, with the appropriate block size to preserve that correlation structure. The VBPBB approach preserves the desired correlation structures within a PC time series while greatly reducing bootstrapping of uncorrelated components.

The VBPBB separates and filters the PC time series by applying a bandpass filter to pass a narrow band of frequencies around the corresponding frequency of the PC component, while frequencies outside this band, the stopband, are attenuated. The Kolmogorov-Zurbenko Fourier Transform (KZFT) filter, an extension of iterated moving averages or Kolmogorov-Zurbenko (KZ) filters, is the band pass filter used for this process and was first developed by Zurbenko (1986). KZ filters and their extensions can separate portions of the frequency domain to exclude interfering frequencies as detailed in Yang and Zurbenko (2010). While other filters may be used, KZFT filters have an extensive history of use in applications in fields including ozone concentration in Tsakiri and Zurbenko (2010), air quality in Kang et al. (2013), global temperature in Ming and Zurbenko (2012), atmospherics in Zurbenko and Potrzeba (2010), climate in Zurbenko and Cyr (2011). Public health applications include research in diabetes in Arndorfer and Zurbenko (2013), and skin cancer in Valachovic and Zurbenko (2014). A frequency separation approach is developed for multivariate analysis on component factors in Valachovic and Zurbenko (2017).

The KZ filter is the iteration of a simple central moving average defined in Zurbenko (1986). This work uses the R statistical software KZFT function in the KZA package. More detail on the KZA package is detailed in Close and Zurbenko (2013). It is a filter with functional arguments $m$, a positive odd integer for the filter window length, and $k$, a positive integer for the number of iterations. KZ filters are symmetric low pass filters, centered at frequency zero, that strongly attenuate signals of frequency $1/m$ and higher while passing lower frequencies, smoothing the time series. The KZFT filter is a band pass filter and has one additional argument $\nu$, and is equivalent to a KZ filter centered at frequency $\nu$. A KZFT filter applied to a random process $\{X(t): t \in T\}$ with three arguments: $m$ time points, $k$ iterations, and shifted center at a frequency $\nu$ is defined in the following equation:

$$KZFT_{m,k,\nu}(X(t)) = \sum_{u=-\frac{k(m-1)}{2}}^{\frac{k(m-1)}{2}} \frac{a_u^{m,k}}{m^k} e^{-i2m\nu u} X(t+u) \qquad (1)$$

The coefficients $a_u^{m,k}$ are the polynomial coefficients from:

$$\sum_{r=0}^{k(m-1)} z^r a_{r-k(m-1)/2}^{m,k} = (1 + z + \cdots + z^{m-1})^k$$

Unlike the PBB approach which only requires the PC component period for bootstrapping, the VBPBB requires the specification of additional arguments affecting the bandpass filters. In this study, for VBPBB to bandpass filter each PC component separately from the COVID-19 mortality time series, VBPBB centers a

KZFT filter at the frequency corresponding to each PC component. Therefore, for each PC component, one KZFT filter has ν set to the PC component frequency. To pass or preserve one and only one PC component in each KZFT bandpass filter, the other arguments set the filter window width of the pass band to be no more than halfway between other frequencies to be filtered. Keep in mind the filter window size, $m$, cannot exceed the number of observations in the time series, and the iterations, $k$, approximately multiply the required observations needed for a given KZFT filter window. Due to the limited number of observations in the COVID-19 mortality dataset, the choice of $m$ and $k$ in this work is restricted and permits little flexibility. Therefore, in this work, for any pair of two PC components with periods $p_1$ and $p_2$ and corresponding frequencies $v_1 = 1/p_1$ and $v_2 = 1/p_2$, KZFT filters centered at $v_1$ and $v_2$ have arguments $k = 1$ and $m$ equal to the closest odd integer larger than $2/|v_1 - v_2| = 2(p_1 p_2)/|p_1 - p_2|$. Generalizing for a time series with $m$ PC components as is this case, where $i = 1, \ldots, m$, for each PC component with period $p_i$ and corresponding frequencies $v_i = 1/p_i$, the KZFT to bandpass that component has $v = v_i$, $k = 1$, and $m$ set to the largest value so that it excludes all other PC component frequencies as described above.

After PC component separation, the VBPBB approach then block bootstraps the PC component time series rather than the original PC time series and uses the same bootstrapping process described above for the PBB approach. This process is repeated for the same large number, $B$, of resamples. Here $B = 10000$, and the 0.975 and 0.025 quantiles of the $B$ bootstrapped time series resamples are calculated to produce a 95% CI for the periodic mean at each time point. This forms a VBPBB 95% CI band for the periodic mean across the time interval.

Significance of a PC component from either PBB or VBPBB methods is determined by what is essentially a horizontal line test for the 95% CI band. If a horizontal line fits through a 95% CI band, then the amplitude of the PC component is not significantly different from zero. A 95% CI band that excludes the possibility of a flat or stationary periodic mean at that frequency indicates a significant PC component at that frequency. Finally, the square of correlation, or coefficient of determination, is used to measure the proportion of variation explained between significant PC components and the original COVID-19 time series.

## 3. Results

The VBPBB 95% CI band for the seasonal mean component of COVID-19 mortality rate ranges from a high in mid-December with a CI of approximately (0.123, 0.203) per 100 thousand to a low in mid-June of (-0.257, -0.120) per 100 thousand people. Figure 2 illustrates the bootstrapped 95% CI band for the seasonal mean variation comparing PBB in red and the VBPBB in blue across typical 365-days seasonal cycles. Across the CI band, the median PBB CI size is 3.49 times larger than the VBPBB CI size. Both VBPBB and PBB 95% CI bands provide sufficient evidence to reject that the seasonal mean amplitude is zero. In other words, the change in the mean that follows a seasonal pattern is statistically significant.

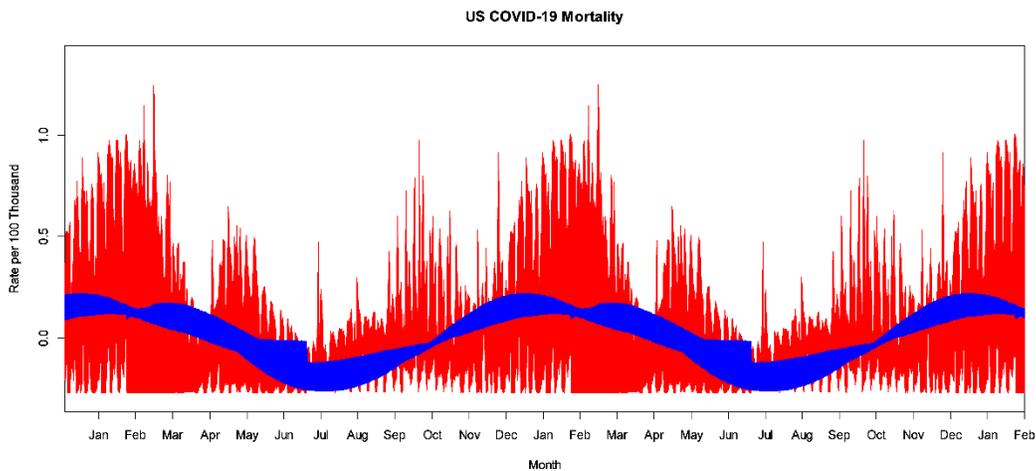

**Figure 2**: 95% CI bands for the US COVID-19 mortality rate 365-day, annual, or seasonal mean variation from PBB in red and VBPBB in blue.

The VBPBB 95% CI band for the second harmonic frequency of the seasonal mean component variation of COVID-19 mortality rate ranges from a high in January and July with a CI of approximately (0.036, 0.129) per 100 thousand to a low in May and November of (-0.126, -0.018) per 100 thousand people. Note that there are two highs and two lows per 365-day period because this is the second harmonic frequency of the seasonal component. Figure 3 illustrates the bootstrapped 95% CI band for the seasonal second harmonic mean comparing PBB in red and the VBPBB in blue across typical half-year cycles. Across the CI band, the median PBB CI size is 5.98 times larger than the VBPBB CI size. The VBPBB 95% CI band provides sufficient evidence that the second harmonic is significant, rejecting that the second harmonic of the seasonal mean variation has zero amplitude, while PBB does not.

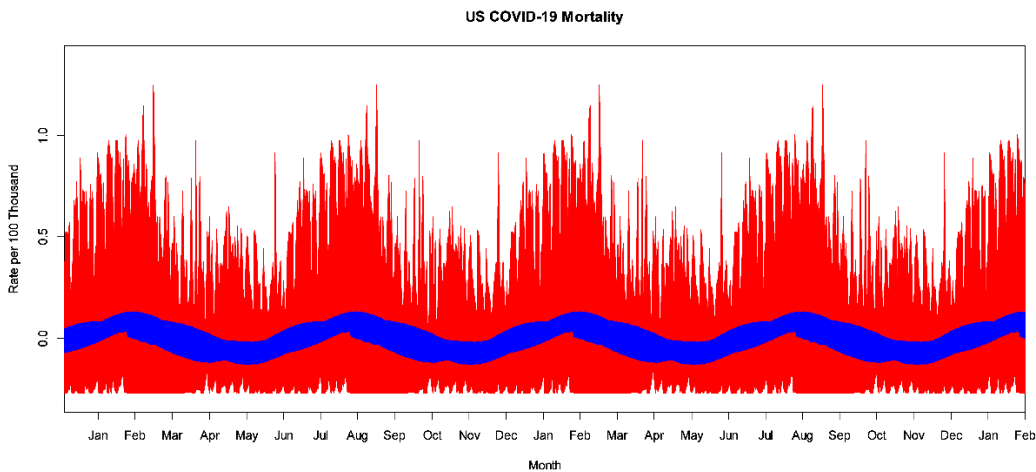

**Figure 3**: 95% CI bands for the US COVID-19 mortality rate seasonal second harmonic mean variation from PBB in red and VBPBB in blue.

The VBPBB 95% CI band for the third harmonic frequency of the seasonal mean component variation of COVID-19 mortality rate ranges from a high in January, May, and September with a CI of approximately (0.009, 0.092) per 100 thousand to a low at the end of March, July, and November of (-0.091, -0.002) per 100 thousand people. Figure 4 illustrates the bootstrapped 95% CI band for the seasonal third harmonic mean variation comparing PBB in red and the VBPBB in blue across typical one-third-year cycles. Across the CI band, the median PBB CI size is 6.41 times larger than the VBPBB CI size. The VBPBB 95% CI band provides sufficient evidence that the third harmonic is significant, rejecting that the third harmonic of the seasonal mean variation has zero amplitude, while PBB does not.

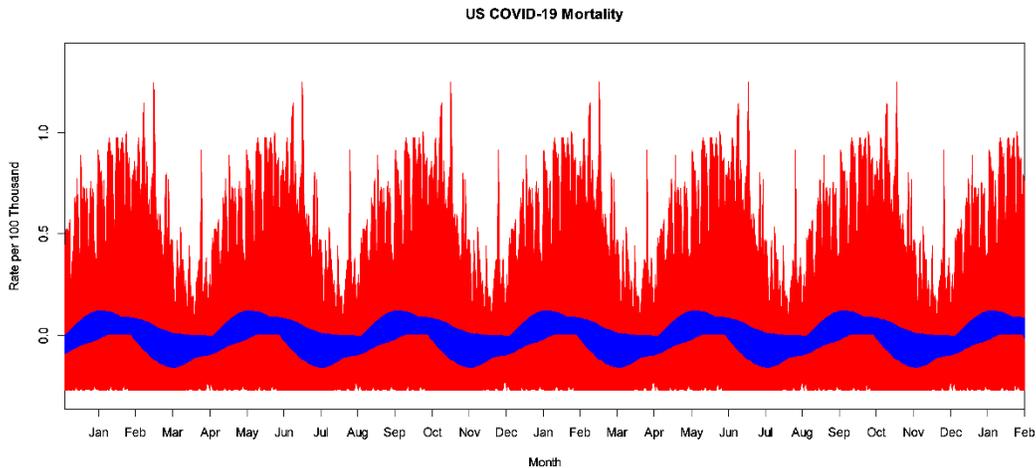

**Figure 4**: 95% CI bands for the US COVID-19 mortality rate seasonal third harmonic mean variation from PBB in red and VBPBB in blue.

The VBPBB 95% CI band for the fourth harmonic frequency of the seasonal mean component variation of COVID-19 mortality rate ranges from a high before and at the start of February, and at one-quarter-year intervals after, with a CI of approximately (0.007, 0.047) per 100 thousand to a low at the start of April, and at one-quarter-year intervals after, of (-0.038, -0.001) per 100 thousand people. Figure 5 illustrates the bootstrapped 95% CI band for the seasonal fourth harmonic mean variation comparing PBB in red and the VBPBB in blue across typical one-quarter-year cycles. Across the CI band, the median PBB CI size is 13.87 times larger than the VBPBB CI size. The VBPBB 95% CI band provides sufficient evidence that the fourth harmonic is significant, rejecting that the fourth harmonic of the seasonal mean variation has zero amplitude, while PBB does not.

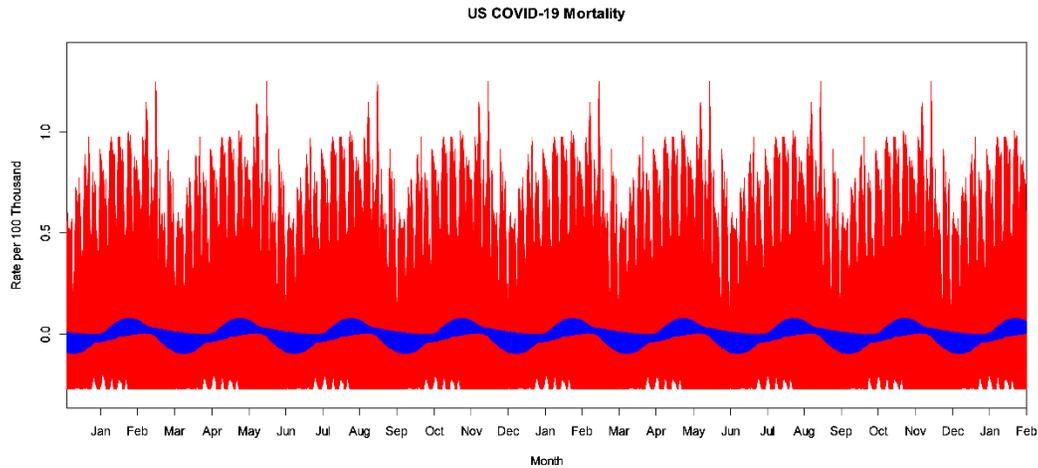

**Figure 5**: 95% CI bands for the US COVID-19 mortality rate seasonal fourth harmonic mean variation from PBB in red and VBPBB in blue.

The VBPBB 95% CI band for the fifth harmonic frequency of the seasonal mean component variation of COVID-19 mortality rate ranges from a high in January, and at one-fifth-year intervals after, with a CI of approximately (0.003, 0.026) per 100 thousand to a low at the start of April, and at one-fifth-year intervals after, of (-0.048, -0.001) per 100 thousand people. Figure 6 illustrates the bootstrapped 95% CI band for the seasonal fifth harmonic mean variation comparing PBB in red and the VBPBB in blue across typical one-fifth-year cycles. Across the CI band, the median PBB CI size is 32.93 times larger than the VBPBB CI size. The VBPBB 95% CI band provides sufficient evidence that the fifth harmonic is significant, rejecting that the fifth harmonic of the seasonal mean variation has zero amplitude, while PBB does not.

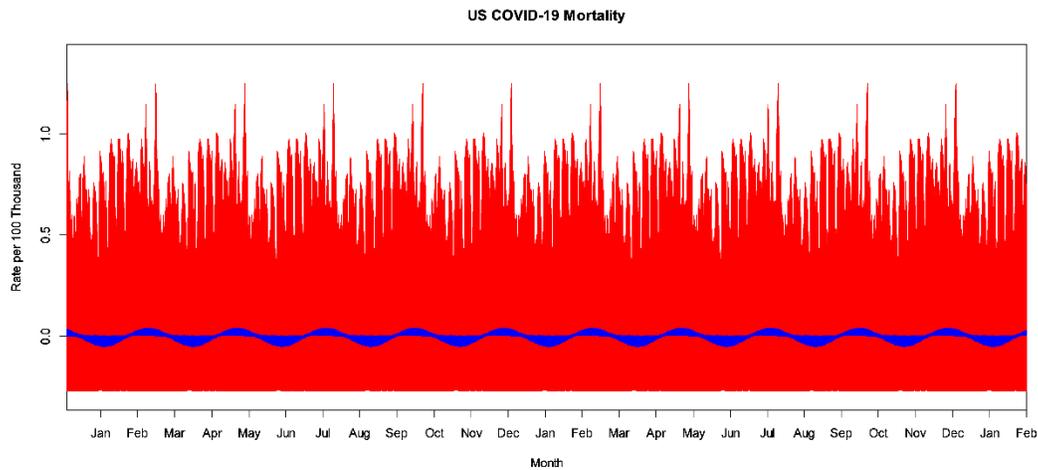

**Figure 6**: 95% CI bands for the US COVID-19 mortality rate seasonal fifth harmonic mean from PBB in red and VBPBB in blue.

VBPBB provides evidence of significant PC components at the seasonal frequency and the second through the fifth harmonics of the seasonal frequency. Analysis of higher harmonic frequencies were insignificant, and results are excluded. The bootstrapped VBPBB 95% CI band for the fundamental seasonal component frequency combined with all significant harmonic frequencies of the seasonal mean component variation of COVID-19 is illustrated in Figure 7 in blue. Combining bootstraps for the CI bands from PBB for the same seasonal harmonics would produce an unreasonably large CI band size as exhibited from the size of the red CI Bands of individual frequencies in the previous figures. Here, the PBB CI band in red is only for the seasonal 365-day mean variation, which we note is still often larger than the VBPBB CI band for all seasonal and harmonic components combined.

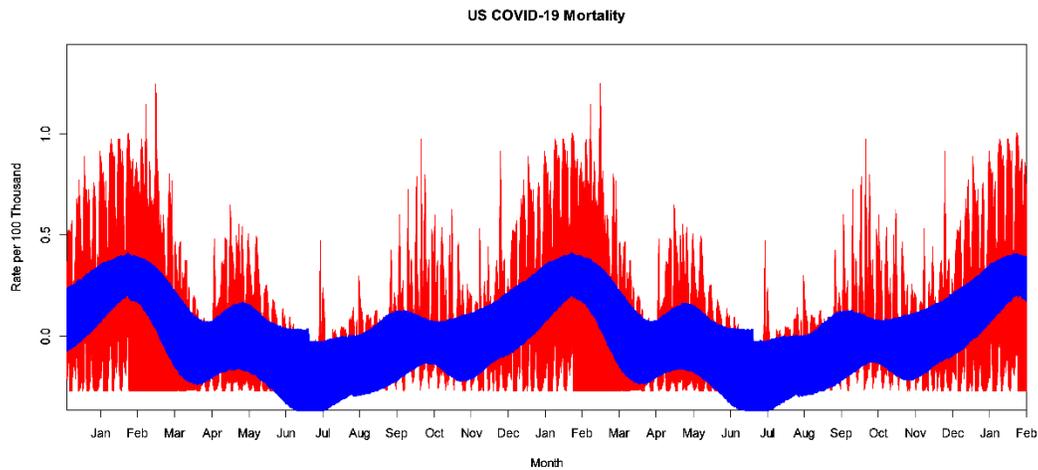

**Figure 7**: The VBPBB 95% CI band for the US COVID-19 mortality rate periodic mean for the seasonal and all significant harmonic components combined is in blue. The 95% CI band for only the US COVID-19 mortality rate seasonal mean variation from PBB is in red.

While interest is primarily focused on the 365-day periodic, or seasonal mean, and associated harmonic frequencies, the VBPBB approach found one additional significant PC component operating at a 7-day period during this research. The VBPBB 95% CI band for the weekly mean component variation of COVID-19 mortality rate ranges from a high over Thursday and Friday, day 4-5 in the weekly cycle with a CI of approximately (0.015, 0.250) per 100 thousand to a low Sunday, day 7 in the cycle of (-0.219, -0.018) per 100 thousand people. Figure 8 illustrates the bootstrapped 95% CI band for the weekly mean comparing PBB in red and the VBPBB in blue across typical 7-day cycles. Across the CI band, the median PBB CI size is 5.47 times larger than the VBPBB CI size. The VBPBB 95% CI band provides sufficient evidence that the weekly component is significant, rejecting that weekly mean variation has zero amplitude, while PBB does not. Additionally, no harmonic frequencies of the weekly component were found to be significant.

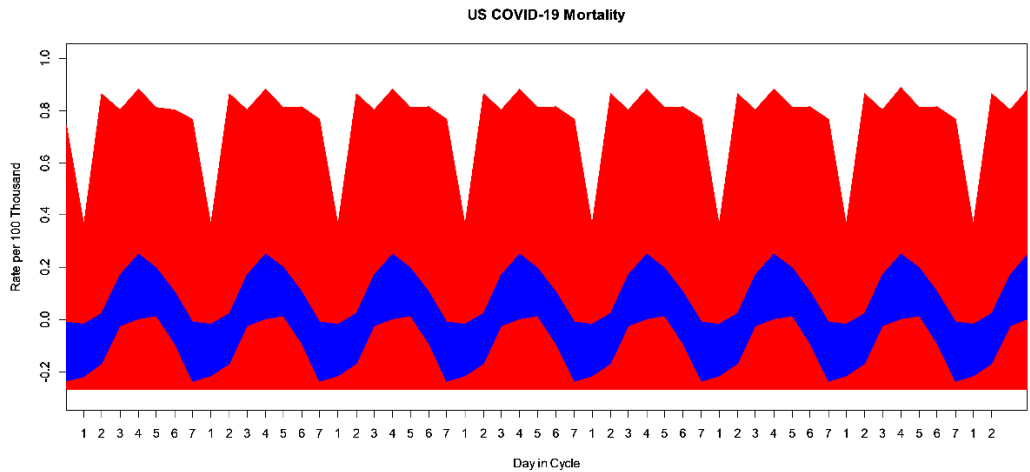

**Figure 8**: 95% CI for the US COVID-19 mortality rate 7-day, or weekly, mean variation from PBB in red and VBPBB in blue.

Finally, the square of correlation, or the coefficient of determination, between the original US COVID-19 mortality rate time series and the median of the significant PC components including the seasonal, seasonal harmonics, and the weekly frequency is 0.186.

## 4. Discussion

VBPBB identified 6 different significant PC components operating at different periods or frequencies in US COVID-19 mortality rates including the season component and its second through fifth harmonics, and the weekly component. PBB was only able to identify one significant PC component, the seasonal component, in US COVID-19 mortality rates. The typical 95% CI, across the CI bands, for the weekly mean from PBB is 5.47 times larger than that from VBPBB and failed to exclude a seasonal mean of zero amplitude. The VBPBB 95% CI band found a significant weekly mean component variation. While this PC component may be due to administrative, systematic, or other human created interventions in documentation and reporting as opposed to natural cycles, it is still critical to investigate and characterize the weekly mean for no other reason that to separate it from an analysis of the seasonal pattern. While PBB is unable to identify this significant component, VBPBB using bandpass filtering of the PC component provides the necessary evidence.

On average the 95% CI band size for the seasonal mean from PBB is 3.49 times larger than that from VBPBB. The 95% CI for the seasonal mean from both PBB and VBPBB excludes a seasonal mean with zero amplitude. Here, both methods provide strong evidence of the significance of the seasonal PC component. However, the general smaller CI sizes across the CI band from VBPBB provides greater evidence of the significance of seasonality, in addition to potential patterns in the timing of COVID-19 waves, and the strength of seasonally affected periods. This is especially true, when the harmonic frequencies of the seasonal component are included to better characterize the pattern. The square of correlation between the median VBPBB bootstrapped seasonal PC component alone and the original time series is approximately 0.092 indicating that the seasonal PC component accounts for approximately 9.2% of the total variability in the US COVID-19 mortality time series. As previously mentioned, the square of correlation, between the original US COVID-19 mortality rate time series and the median of the significant PC components including the seasonal, seasonal harmonics, and the weekly frequency is 0.186 implying that the median of the statistically significant VBPBB bootstrapped PC components explains 18.6% of the variability in US COVID-19 mortality rate.

Only VBPBB provided evidence for the significance of the second through the fifth harmonics of the seasonal PC component. PBB, with much wider CI bands, was unable to exclude a periodic mean variation for these components as having zero amplitude, and insignificance at these frequencies. On average the 95% CI band size for these harmonic means from PBB ranged from 5.98 to almost 33 times larger than that from VBPBB. The VBPBB 95% CI band at peak and trough times for each PC component can be found in the results section, but the VBPBB bootstrapped combination of the seasonal and seasonal harmonic frequencies provides a useful picture of the seasonal pattern, seen in Figure 7. Across the VBPBB CI band we notice relatively smaller crests in US COVID-19 mortality at the end of April and the beginning of September, and a large periodic crest at the end of January. The periodic variation in US COVID-19 mortality has troughs in March and October, and the most pronounced trough in June. Compared to PBB, VBPBB provides a powerful tool to characterize daily confidence intervals as well as annual CI bands for the mean variation occurring as a result of the combined effect at these significant PC components. This should prove beneficial for COVID-19 public health planning.

VBPBB has several limitations in its design in comparison to other PBB methods. VBPBB performance is tied to selection of arguments used for bandpass filtration and results may differ depending on the choice for PC component estimation for the US COVID-19 data. In fact, since VBPBB bandpass filters the original time series prior to block bootstrapping, PBB is equivalent to a trivial case of VBPBB. PBB is a type of VBPBB when the bandpass filter applied is so wide that it passes all frequencies (equivalent to $m = 1$, a moving average of one, in the KZFT filter), attenuating no frequencies prior to block bootstrapping. Given the improvement and better performance of VBPBB to PBB, it is reasonable to ask if different argument selection for the bandpass filters would affect performance and results, and if they could be improved further. These are questions for future work. Also, since VBPBB methods rely upon filtering the original PC time series with the application of moving averages, the observations at the ends of the time series will have the filter incompletely applied. For this reason, VBPBB performs better with greater quantities of data than PBB, and results will likely improve with additional COVID-19 data.

Bootstrapping in general is a method well suited for limited assumptions and relatively little data, but there are limits. Currently, with between three and four years of COVID-19 data, bootstrapping methods are just beginning to be useful for seasonal characteristics. Therefore, these results should be viewed as preliminary. Also, this work should not be generalized to any other groups or populations other than the collective national population. However, it is illustrative of the existence of seasonality and other components in COVID-19 nationally, and demonstrates the potential to extend this VBPBB analysis for stratification by state, county, hospital, demographics, etc. We recommend the use of this novel VBPBB method for additional COVID-19 datasets, including test positivity, ER reported cases, hospitalizations, and other metrics, all of which can aide public health preparation and response. As additional data becomes available, VBPBB performance is likely to improve, and results likely to become more reliable. Additional PC components may become significant with additional data, and existing CI bands for the PC components of COVID-19 mortality revealing timing and intensity will be refined. We recommend that additional research continue particularly because of the ongoing concern and importance of characterizing COVID-19 seasonality.